# Understanding misinformation in India: The case for a meaningful regulatory approach for social media platforms

GANDHARV DHRUV MADAN, 25, BBA+MBA (IIM INDORE), BUSINESS MARKETING (HCL TECHNOLOGIES)

i15gandharvd@iimidr.ac.in, +91-9650022633

**EXECUTIVE SUMMARY**

This paper will examine misinformation in India – understanding its nature, the context in which it thrives and a brief note on the real-world consequences that are prompting discourses like this paper. Consequences and context, being the major themes driving this research effort. Why India as a society is a hotspot for misinformation, how social media platforms have failed the general public in their actions/inactions, and the business and sociological implications of the same. Covering the actions from platforms to mitigate harmful content, while understanding and rationalizing the options for future steps in this direction. On the government's side, focus on the issue of changing/diluting the intermediary liability regime and the IT Act, which are currently ill-equipped to address online misinformation while threatening the democratic rights of free speech and privacy. The paper builds the argument for platforms to see regulation in their benefit and avoiding the typical capitalistic consequence of the 'tragedy of commons', along with the government to see platforms as an ally in their initiative to achieve sustainable and accountable public good. The paper makes its stand clear on the position and scope of government intervention that is required to address this situation by reinforcing existing research on regulation codes and social media practices. This paper makes a strong business and sociological argument for a joint public-private exercise in spearheading social media 'self-regulation'. A partnership that will be anchored in democratic virtues of freedom of speech, active citizenry, etc., and driven in the interest of the industry for a technology-backed, future-ready process of redefining social media commerce.

**THE MISINFORMATION PROBLEM**

India is a misinformation-prone South Asian country. Although political misinformation is commonplace in India, it has often been overshadowed, or even accentuated, at times, by health and religious misinformation. Political misinformation only surges before or during important political events. [1] The existing political environment is responsible for political and other categories of misinformation. Two political conditions are usually conducive to misinformation: polarization of the society and populist communication. Both tendencies are present in contemporary India. In addition, low trust in news, weak public service media, a more fragmented audience, and high social media use can also be responsible factors for misinformation in India. [2] Interestingly, instead of taking a comprehensive and creative anti-misinformation initiative, the Indian government shuts down internet services quite often (95 times in 2019) [3] to cease misinformation. [4]

Moreover, the country's law makes tracing the rumour-spreaders difficult. [5] Though a few digital interventions, such as chatbots and Aarogya Setu, have been seen during the pandemic to combat the proliferating misinformation problem, these are creating a crisis of trust among people in different ways. [6] India as a democracy has had a mixed and chaotic track record in combatting fake news. Although the governments have shown to be continuously working with big technology companies to restrict the spread of fake news, some of its politicians have continued to undermine the government's effectiveness by spreading unproven and inaccurate health information (Mohan, 2020). However, the Indian central and state governments have continued the strategy of the Internet shutdown in various parts of the country during COVID-19 to control the flow of information, which some argue is more to do with controlling political opposition and civil protests (Nazmi, 2019).

**LITERATURE REVIEW & METHODOLOGY**

For research, this paper has included numerous literature that are covering a variety of information on the topics of misinformation, social media and fake news, regulation of misinformation and social media platforms, all presented for India. Studies including thematic analysis of misinformation, brief history on social media and its amplification of misinformation, current and past policy interventions by the Indian government, history of self-regulations in industries, and an analysis of regulatory approaches in the Indian context. This paper aims at introducing a coherent reading into the context of misinformation in the country and the subsequent social and business disruptions that will follow. Utilizing lessons from history around industry regulations, existing policy research and framework analysis to convince the reader of the nature of policy intervention that will bode well for all stakeholders involved. The literature sources have been mentioned in their respective sections for reference.

The research utilized the PASTEL framework to analyse data collected from other research efforts covering the topic of misinformation and regulation across academic whitepapers and news media blogs and articles, all available freely on the public domain. Relevant secondary data, in terms of information, previous analysis in other research efforts, and literature work included in respective sections in the paper have been reproduced, shared and/or indicated wherever necessary.

**MISINFORMATION AND SOCIAL MEDIA**

About 86% of 1.38 billion Indians subscribed to a mobile phone at the end of January 2020 (TRAI.gov.in, 2020). An overwhelming majority access the wireless service via private providers, while about 10% of mobile users subscribe to a public service provider. Most of the Internet users (687 million) access the Internet and news on their mobiles (Statista.com, 2020a). With over 400 million monthly active users, the social networking site Facebook dominates the Indian market, while WhatsApp is the most popular messenger application (Datareportal.com, 2020).

Efforts towards mass media literacy and misinformation awareness, like the 2020 documentary, "The Social Dilemma", [13] and organizations, such as the Centre for Humane Technology, [14] raise a number of legitimate concerns about the negative impacts of social media technology on our society. Media technology has reached a point where it's influencing social behavior. The internet and social media have decentralized and democratized content production and distribution more than ever. Similarly, the reach and speed of the internet is far more accessible than before. Internet media companies (social media platforms, OTT platforms, etc.) use algorithms to take advantage of concepts from psychology (like variable ratio reinforcement, the reward pattern of slot machines) to increase the addictiveness of their software. While media and advertising have long made use of these tactics, the internet now enables rapid improvement of the best attention-grabbing techniques via the standard practice of massive scale A/B experimentation. Mobile software leverages push notifications to grab your attention even when you have not explicitly chosen you want to look at your phone. This has made one of the most distasteful elements of consumer capitalism – advertising – more pervasive than ever. We can no longer call most of the spaces we occupy outside of our home or office "public" spaces – they are nearly all "commercial" spaces. The internet is far more a commercial space than it is a public space. Social media leverages friend, or 'authority' endorsements to earn your attention and trust for their information. You are more likely to click / trust a link accompanied by a face you know. This is an age-old tactic of making appeals to the authority of tribe, of trying to persuade you by showing you "people just like you" rather than making an appeal to reason. A common practise across Indian politics and society as well. [15]

Social media is using these attention-techniques, common in advertising and public relation exercises through existing consumer technologies to create the perfect nightmare scenario of a polarised society, captured within their respective information echo chambers, living by divisive rhetoric and ideologies that are harmful to social unity and the ideals of democracy. In past generations, a household consumed the same content (a single copy of a newspaper), often together, at the same time (sitting around a radio or TV together). Mass market paperbacks and cheap consumer electronics – VHS, cable TV, personal devices like the Walkman, etc – arguably were all steps towards "personalized" (and isolated) consumption. While fragmentation is not necessarily a totally new phenomenon, it is fair to say that never before has every individual's world been so personalized, fragmented, and disconnected. This cannot be good for social cohesion. [15] Addiction to mobile devices can result in isolation and depression. Demagogues and profiteers can exploit powerful new technology to amplify their own voices for personal gain. Misinformation and 'fake news' can degrade trust in institutions and fragment our worldviews with conflicting ideas about what constitutes reality, making it difficult to reach consensus on public policy. The more divisive we become as a society, the more radical representation we will seek in our legislative bodies, effectively destroying the concept of common ground in the parliamentary democracy of our country, where we pride ourselves on the principles of 'unity in diversity'.

While we're identifying many of the systemic failures that have been introduced (or exacerbated) by technology. Social media sites and apps with their capacity to allow many-to-many communication and the ease of sharing of micro-content have become an important avenue for exchange of news and views in recent times. However, the proliferation of fake news, the concerns about data privacy and the commercialisation of user data have ensured that the debate about government regulation of social media platforms continues. Social media companies are coming under increased pressure to moderate misleading and false content being shared on their platforms. Scholar have noted the changing discourse about social media being a tool for democratic uprising (Lewis and Molyneux, 2018) and activism (Xu, 2016) to being a means for government surveillance, corporate data collections (Vaidhyanathan, 2018), political manipulation (Gunther et al., 2018; Rodrigues, 2019) and a space for radicalisation (Figoureux and Van Gorp, 2020). Governments in more liberal and democratic countries, with higher level of media and informational literacy, remain torn between the need to preserve citizens' freedom of speech and curtail the circulation of fake news (Rochefort, 2020).

Meanwhile, India also has one of the largest traditional media with over 100,000 news publications, about 850 television and 1000 radio channels. The media derives its freedom, to question those in power, from the Constitution of India's article 19, which guarantees all its citizens the fundamental right of 'Freedom of Speech and Expression', with a few caveats such as free speech cannot be used to promote communal disharmony, go against public decency, and compromise security and national sovereignty. However, on the 2020 World Press Freedom index, India dropped to 142nd ranking because of recent attacks on journalists by pro-government political activists and the imposition of electronic and the Internet blackouts in the Kashmir Valley (RSF.org, 2020).

**EXISTING ACTIONS AND PUBLIC POLICY INTERVENTION**

<u>THE GOVERNMENT</u>

Government efforts to address misinformation faces a number of challenges including jurisdictional tensions and the complex nature of imposing new rules on freedom of expression. Vague, top-down rules, such as the new Intermediary Liability Rules, which force platforms to police content or prejudice users' freedom of speech, are not only harmful to users but also potentially unconstitutional. [17] However, the Indian government has come up with a new set of rules and regulations to regulate social media platforms, messaging services, OTT platforms and news portals. These regulations are called the Information Technology (Guidelines for Intermediaries and Digital Media Ethics Code) Rules, 2021 (Rules). [18] These Rules will require compliance even by foreign tech-giants operating in India such as WhatsApp, Facebook, Twitter, Netflix, Amazon etc. For social media platforms such as Twitter, Facebook and others, the rules focus on issues like fake news, fake user accounts, originator of messages, and monitoring of illegal content by platforms. Greater compliances are placed on social media platforms with larger user base. Social media platforms with minimum 50 lakh registered users are classified as significant social media intermediaries and are subject to maximum compliances. However, the Government may require any other social media platform to also comply with rules applicable to significant social media intermediaries if services of such platform create a material risk to the sovereignty or integrity of India.

All significant social media intermediaries are required to appoint, a Chief Compliance Officer, Nodal Contact Person and, a Resident Grievance Officer. Each of the above are required to be Indian residents. The Rules also necessitate significant social media intermediaries to have a physical contact address in India. This mandatory physical presence in India will have significant implications for foreign players in terms of setting up infrastructure and deployment of resources and taxation. Significant social media intermediaries are now required to deploy technology-based measures, maintenance of appropriate human oversight, and periodic review of automated tools. Such active monitoring by intermediaries dilutes the safe harbour protection that was available to intermediaries under the 2011 Rules. [18]

<u>SOCIAL MEDIA PLATFORMS</u>

Facebook, Twitter, WhatsApp, and Google, among other social media and messaging platforms, have reportedly focused on bringing transparency to aspects of political content, verifying political advertisers, providing transparency in expenditure on political advertisements, more rigorously monitoring content on their platforms, responding to government requests, including from the Election Commission of India, [19] and raising awareness about political content. For the elections, Facebook has also been banning false accounts and collaborating with third-party fact-checkers. While platforms have undertaken awareness campaigns and partnered with local law enforcement officials and policymakers to combat fake news, such measures are often ad-hoc, and not tailored to unique challenges faced by the Indian user base.[20] Many concerns have been raised regarding the efficacy of existing fact-checking programs. For instance, Facebook's third-party fact checking program was rolled out in Karnataka in 2018 and later expanded to cover other Indian States.[21] More recently, it launched a partnership with independent entities to identify misinformation across 11 Indian languages.[22] However, some experts have expressed concern over the inefficiency of

Facebook's partner fact-checking organizations [23] and the lack of action taken by the platform to remove flagged content, rendering the fact-checking process ineffective. [24]

Likewise, WhatsApp has introduced voluntary third-party fact-checking systems [25] and restrictions on bulk-messaging [26] to make it tedious for users to forward content to hundreds of users in one go, in order to dampen the virality of harmful content. However, it has been argued that these changes are ineffective and superficial. [27] For instance, the restrictions on bulk-messaging can be bypassed through affordable "clone applications" used by political party workers to forward content to a large number of people, or the use of anonymous phone numbers to send bulk messages. [28] WhatsApp has also "launched a fact-checking hotline" that encourages users to flag messages for verification. "Those monitoring social media content believe actions taken by social media giants to limit fake news are not even scraping the surface of the problem, since fraudulent posts are disseminated quicker than they can be taken down." [29]

POLICY INTERVENTION [30]

A way out might be to retain the intermediary construct, but with more clear-cut obligations for platforms to qualify as intermediaries and enjoy safe harbour protection. This may be achieved with the following steps, carried out through amendments to the IT Act:

**Define intermediaries:** The definition of intermediaries needs to be updated, rationalised and subdivided according to functionality, such as: social media platforms, private messaging platforms, search engines, web-hosting services, and internet service providers. Obligations for each of these sub-categories may need to be suitably tailor-made, given their distinct modes of operation.
**List banned content:** The types of content that are deemed to be harmful and illegal should be specifically and exhaustively set out. This is required as the IT Act presently does not adequately address content that can adversely affect individuals at a personal level, such as online threats, harassment and breach of privacy. It would also help address challenges such as hate speech, fake news, information and psychological warfare.
**Lay out govt's powers:** The powers of the government with respect to illegal content should be set out. This could include the power and procedure to take down content, situations where decryption or interception of communications is necessary, and appropriate punishments for certain forms of content.
**Minimise platforms' takedown powers:** Suo-moto content takedown powers of platforms should be kept to a minimum in the case of verified users – limited to things like pornographic and child sexual abuse content. This would prevent platforms from becoming laws onto themselves, with a free hand in controlling the discourse on their platforms.
**Discourage echo chambers:** Platforms should be encouraged to do away with algorithms that promote certain types of content by way of user profiling and selective highlighting. This will help prevent platforms from becoming information echo chambers that reinforce polarisation and prejudices.

**SELF-REGULATION**

Digital platforms are highly profitable businesses that have also enabled the distribution of fake news and fake products, manipulation of digital content for political purposes, and promotion of dangerous misinformation on elections, vaccines, and other public health matters. [31]

The 'social dilemma' is clear: Digital platforms can be used for evil as well as good.

As platforms emerge as "modern public squares" [32] and make speech decisions for billions of people online, there is a need for efficient regulation of the ecosystem in which misinformation thrives. Thus far, neither platforms nor the government have effectively performed this function. This research argues for a new participative and responsive rule-making structure, and an accountable and transparent platform governance framework for India.

Governments will inevitably get more engaged in oversight. However, we believe that platforms should become more aggressive at self-regulation now. To explore the feasibility of self-regulation, we found that companies have often risked creating a "tragedy of the commons" when they put their short-term, individual self-interests ahead of the good of the consuming public or the industry overall, and, in the long term, destroy the environment that made them successful in the first place.

History provides several lessons for today's digital platforms.

For many decades, companies in the business of producing movies, video games, and television shows and commercials all have faced issues around the appropriateness of "content" in a way that resembles today's social media platforms. To keep regulators at bay, the movie and video games industries resorted to a self-imposed and self-monitored rating system, still in operation today. The broadcasting and advertisement sectors in the 1950s and 1960s faced pushback on the appropriateness of advertisements, with issues resembling what we see today in online advertising.

Self-regulation in these cases often delivered effective and inexpensive guidelines for company operations as well as forestalled more intrusive government intervention. At the same time, these historical examples suggest that self-regulation worked best when there were credible threats of government regulation. The bottom line: Self-regulation may be the key to avoiding a potential tragedy of the common's scenario for digital platforms. [31]

Second, we find that firms in new industries tend to eschew self-regulation when the perceived costs imply a significant reduction in revenues or profits. Managers rarely like industry regulations that appear "bad for business." However, this strategy can be self-defeating. If bad behavior undermines consumer trust, then digital platforms will not continue to thrive. These 'online intermediaries' have broad immunity from liability for user-generated content posted on their sites. They generally resisted and argued that their legal and political positions would be more secure if they avoided potentially controversial curation. Internal debates ranging from free speech versus censorship to how much curation can the firm perform before it crosses the line from platform to "publisher" led most social media companies to resist aggressive curation until very recently.

Third, proactive self-regulation was often more successful when coalitions of firms in the same sector worked together. We saw this coalition-type of activity in movie and video-game rating systems limiting violent, profane, or sexual content; television advertisements rules curbing unhealthy products like alcohol and tobacco; and computerized online airline reservations giving equal treatment to airlines, without favouring the system owners. Similarly, social media companies implemented codes of conduct on terrorist activity. Since individual firms may hesitate to enact self-regulation if they incur added costs that their competitors do not, industry coalitions have the benefit of reducing free-riding. Now is the ideal time for more "coopetition," where platforms compete as well as cooperate with rivals.

In sum, history suggests that modern digital platforms should not wait for governments to impose controls; they should act decisively and pro-actively now. While the costs of government action in the internet era have been modest so far, the regulatory environment is changing fast. Going forward, governments and digital platforms will also need to work together more closely. Since more government oversight over Twitter, Facebook, Google, Amazon, and other platforms seems inevitable, new institutional mechanisms for more participative forms of regulation may be critical to their long-term survival and success. [31]

In India, platform responses to misinformation have been inconsistently applied across regions and political groups, and suffer from a lack of transparency. Misinformation challenges are continuously evolving and, as a result, solutions are difficult to implement. However, technological changes that have been introduced thus far have been inadequate towards the fundamental goal of empowering users to identify fake news and restrict its circulation. Platforms have also failed to engage with communities and civil society in collaboratively addressing the misinformation challenge in India. More broadly, platforms are not adequately accountable to State and local governments, as exemplified by Facebook's refusal to appear before a legislative committee investigating the platform's role in riots in India's capital. [32]

At present, self-regulation by platforms has failed to generate satisfactory solutions to address misinformation in India. There are a few instances where such efforts have been initiated, but their functioning is largely opaque. The Information Trust Alliance ('ITA') is one example of a recent collaborative effort by platforms, digital publishers, industry bodies, fact-checking organizations, civil society and academics. [33] The aim of the ITA is to work together to develop standardized procedures for resolving complaints regarding false or disputed content. [34] The ITA has reportedly drafted a Code of Practice, which is on hold due to a lack of consensus among participating social media platforms. [35] As of early 2022, little is known about the deliberations of the ITA beyond this.

In order to address the misinformation challenge in India, rather than prescriptive intermediary liability laws, Indian lawmakers should work with platforms to develop self-regulation or co-regulation that is responsive and collaborative, while still sufficiently accountable to domestic policymakers. This is where "induced" self-regulation [36] or "regulated" self-regulation [37] may be an approach worth considering. In its most basic form, "induced" self-regulation refers to the creation of an enabling regulatory environment which incentivizes or induces industry actors to create rules for self-regulation, while also guarding against state interference or a compromise of the independence of entities. [38] It creates a controlled framework in which industry actors have flexibility to devise norms, but at the same time must be accountable for the enforcement of such norms.

**POLICY ANALYSIS: PASTEL FRAMEWORK** (Table 4 and 5)

**Do Nothing:**

Misinformation continues to thrive, furthering divisive rhetoric across political, religious and geographical lines with no repercussion for the ensuing social disruption. This clearly will fall on either side of political feasibility, due to the current nature of the ruling party and its oppositions' differences. This will have a negative impact on administrative and social feasibility due to the reasons mentioned above. A similar impact will be observed on the legal aspect, due to the increasingly intrusive nature of information and advertising having no legislative action to maintain control. Economically, this will continue to benefit the platforms, with no regulations, they can continue to reap in audience attention and the subsequent ad revenues. Which, in turn, will continue to take viewership and consumer mind space away from the traditional news networks, as seen from the ongoing transition in our thematic analysis. No impact technologically with this inaction.

**Platform-controlled regulation:**

With complete flexibility on control, it is this paper's perspective that the situation will remain similar to the 'Do nothing' option, unless there are caveats around public disturbances and information manipulation inserted in the regulation. Giving platforms the power to control content dilutes their intermediary status, which in turn, will allow for criminal liability to be charged in certain circumstances. Effectively, this solution doesn't seem favourable to any of the stakeholder's concerned. Economically, this will continue to benefit the platforms and advertisers. Apart from that this is not politically, socially or legally feasible, due to the reasons mentioned above. Administratively, this can go either way, with the government either being viewed as progressive, or not doing its part for the citizens. For political feasibility, this might place the governments at the mercy of the platforms for disseminating their message, something political parties have taken full advantage of in modern electoral history. Falling out of favour with the platforms may not be in the government's interest, basically defeating this option for any consideration altogether. No impact technologically with this inaction.

**Government-controlled regulation:**

Government taking full control of content regulation on social media platforms might effectively render them as public new media intermediaries, which may not be a negative, however, our analysis has to be rooted in current realities. Therefore, with governments and political parties being extremely protective about public discourses, especially when it sheds any non-positive attention on the incumbent administration, platforms, advertisers and the public might suffer the most in this scenario. Although this move might be feasible politically and legally, assuming full control completely removes any content liabilities from the platform, this will have a negative impact socially, economically and technologically, as is the case when slow-moving public administration gets involved in stringent regulations. Administratively, it would be a toss-up between carrying out the content censorship and how effective and/or ineffective the performance metrics would turn out to be, the public perception fallout globally notwithstanding. The industry might find another way to innovate content delivery eventually, maybe a decentralized social network, or just shut shop.

**Joint Self-regulation:**

This option presents a rather ambiguous middle-of-the-road solution, unless effective participation, collaboration, and liability frameworks are drawn up for both platforms and the government. As has been discussed in the paper, induced regulation is a 'common good' solution where the platforms get to enjoy their intermediary rights, as well as act as responsible entities in a national context, with the support of the administration. The development and oversight of induced self-regulation or co-regulation should be undertaken by an independent body with adequate tools of enforcement and oversight. This option holds positive impacts for all stakeholders, with an objective assessment of finding situations where the government or advertisement firms might end up on the wrong end of a censorship conflict with the platforms. It holds a positive impact across all feasibility options, with only the potential of legal issues getting in the way of the actions and its rationale taken by the platforms. This solution, apart from its social positives, holds a business rationale, with the avoidance of strict industrial regulation, and the creation of new use-cases for internet technologies and compliance/regulatory functions.

**Independent institutional regulation:**

A solution that holds some historical rationale as well. Public policy rarely ever manages to keep up with disruptive trends and industries in terms of protecting citizen's rights and promoting commerce, the cryptocurrency regulations being a prime example. Simultaneously, private enterprises cannot be expected to act in good faith when it comes to situations that are against its business interests. Enterprises are answerable to their shareholders before addressing any kinds of moral liabilities/dilemmas, take any enterprise in the energy industry and climate change for reference. At least on paper, establishing a quasi-judicial body set up as a nodal tribunal for all actions taken by the government and platforms on user content, and for hearing users' grievances against such actions does sound promising. What might be considered a 'make it or break it' in this case would be how much regulatory power is given with the institution, what are the liability and stakeholder engagement frameworks in effect, etc. This institution could be a great symbol of civil engagement to solve a global problem. However, there are considerations that need to be addressed, both from the government's perspective and the platforms', to ensure an effective regulatory approach, that not only solves political and civil concerns, but manages to do so while keeping business interests in mind. History has also shown how these independent institutions end up failing with their tall task orders without the compliance of every party involved. The tribunal will be toothless without the technology wherewithal of the platforms, and the law enforcement of the government. Especially in a country like India, where the government is used to enforcing their will on independent institutions, and global media platforms, that have shown their refusal to entertain local law enforcement in the matters of misinformation. This paper doesn't reject this alternative altogether, but from the literature and data analysis in this research, it definitely feels like the second-best alternative in this context. In its recommendation, the paper does agree that there is a requirement for an institutional arrangement in the development and oversight, with sufficient power of enforcement, of the regulatory code that can support the government and platforms in combatting misinformation. When it comes to addressing its stakeholders and impact on various kinds of feasibilities, the above-mentioned fact becomes evident. This might be a scenario which doesn't carry real-world potential currently.

**POLICY INTERVENTION AND RECOMMENDATION**

This paper reproduces the recommendations of another paper with further historical evidence and original analysis to propose India to consider a voluntary code for misinformation, including key norms for social media platforms to adhere to. Such a code should outline outcomes for platforms to achieve, including design-duties [39] and product features to empower users and fight misinformation. [40] The standards should be developed collaboratively through transparent and participative processes, and governments should evolve efficient metrics for assessing their performance. The outcomes should be grounded in human rights principles and developed through a collaborative process involving the platforms, civil society, and other stakeholders. Some examples may include protecting the free speech and privacy of users, bolstering user autonomy and access to remedial measures and grievance redressal, and greater transparency and accountability around content moderation and efforts to combat misinformation. The outcomes should be built around common objectives, and should provide flexibility for platforms to develop protocols and technological tools to achieve them.

Evolving practice globally also indicates a greater push towards accountable and collaborative self-regulation to address misinformation. For instance, the Australian Communications and Media Authority ('ACMA') has recommended a self-regulatory approach [41] to enable platforms to identify risks or "acute harms from misinformation" [42] and devise their own design solutions to empower users. The outcomes outlined by ACMA include reduced exposure of users to harmful misinformation on the platform, a robust system of misinformation reporting, and access to an effective "complaints handling process". [43] The proposal argues for the need to maintain consistency in measures to tackle misinformation, using "industry-wide" inputs for risk-assessment. [44] It also seeks to push platforms to develop common objectives to curb misinformation, [45] but give them the flexibility to develop their own solutions to achieve the objectives, [46] and provide a system of "performance reporting". [47] This mechanism aligns with the "positive state approach" where governments create institutions with incentives for private actors to tackle a challenge, rather than adopting a system which relies on fines and penalties. [48] This may be contrasted against a "negative state" approach in countries like Germany [49] and Singapore, [50] where restrictive regulation poses a threat to the free speech of users.

In India, this might be carried out by an independent platform oversight body which operates at an arm's length from both the government as well as platforms. The platform oversight body should ensure that the development of these codes is participative and promotes dialogue among state and local governments, platforms, civil society, academics, and other stakeholders. [51] The codes should outline key performance indicators and lay down a metric for the independent body to monitor compliance. In order to ensure accountability and oversight, codes which are independently developed by platforms may be approved or accredited by the oversight body.

Since India does not have an existing legal framework around platform governance, an outcomes-based self-regulatory framework can serve as a necessary first step for stakeholders to work together to combat online misinformation. In the absence of enforcement tools, its implementation hinges on platforms' willingness to work together. A potential oversight body may also face hiccups in terms of conducting audits or effectively monitoring implementation. However, this approach would be useful in bridging the gap between the government and social media platforms, and potentially avoid prescriptive regulations such as the amendments to the Intermediary Liability Rules which pose a larger threat to user rights. In the short-term this framework should, at the very least, prod platforms to take holistic action in curbing misinformation and creating an environment of greater transparency around their actions in this space. [52]

**CONCLUSION**

The conversation around misinformation on social media is evolving to understand the dangerous consequences of non-action by the platforms and the government. Regulation of any sort, developing a code, the oversight and compliance structure, enforcement authority and collaboration frameworks, will require a change in current mindset from all ends of the discourse. This paper has so far outlined critical areas of reform which can be prioritized in the regulatory effort. Beyond the public policy aspect, the research also covered the social and financial harm that should be expected, at times, already being experienced, when misinformation is allowed to thrive in the attention economy of our world. As all modern policy solutions go, this will be an iterative process, requiring community engagement, a state-platform consensus and the will to alter the regulatory landscape. It is hoped that the analysis and recommendations outlined in this paper will help in furthering the exploration of regulatory actions for India, and assist similarly situated countries in combating misinformation on social media.

**APPENDICES**

ANALYSIS 1: MISINFORMATION THEMES

The word "misinformation" becomes more popular among communication theorists after the 2016 US election. [7] Though four terms—misinformation (unintentionally misleading information), disinformation (intentionally misleading information), rumor (unverified and/or misleading information), and fake news (intentionally misleading information that can be true or false) are distinctive if we consider it from the orthodox philosophical viewpoint. Also, the term "misinformation" is used conveniently to denote misleading information whether it is intentional or not, and verified or not; [8]; [9]; [10]; [11], [12]. This study follows a similar conceptual footprint as well, using "misinformation" as a representative term. While misinformation is an old phenomenon, it becomes a modern-day problem around the world thanks to the Internet and social media.

Reproducing the thematic study conducted to analyze Indian social media fake news. [16] The specific focuses of this research were to identify the main themes, content types, and sources of fake news. An analysis of 419 social media fake news collected from an Indian fact-checking website produced some novel findings. This paper is sharing the results as the analysis is helpful to the goal of getting a meaningful understanding of the nature of social media regulation that should be recommended to policy makers.

The results show that fake news on social media has six dominant themes: health, religion, politics, crime, entertainment, and miscellaneous (**Table 1**). Health-related fake news is on the top of the list with a frequency of 114 (27.2%), followed by fake news about religion (n=105; 25.1%) and political fake news (n=102; 24.3%). These three themes make up 76.6% of the fake news stories in this study. In religious fake news, text & video has the highest share (n=44; 41.9%), followed by text & photo (n=39; 37.1%), whereas audio and video have the lowest shares (both n=1 and 1%) (Table 3).

Stories that include text & photo made up the largest share of fake news stories about politics, miscellaneous, health, and entertainment, 39.2% (n=40), 43.8% (n=14), 36.8% (n=42), and 61.9% (n=13) of stories, respectively. In crime-related fake news, text & video has the highest share (n=18; 40%). Of the six themes, crime-related fake news has the highest percentage (97.8%) in online media, followed by health (93%) and religious fake news (87.6%), while entertainment fake news is the lowest (66.7%).

Fake news in social media can take eight forms: text, photo, audio, video, text & photo, text & video, photo & video, and text & photo & video (**Table 2**). While the first four are the primary content, the other four are combinations of one or more primary content. These combination categories are needed because a single piece of fake news can be found in two or more forms at a specific time. In this typology, text & photo appears more often than the others with 165 of the stories analyzed (39.4%), followed by text & video (n=126; 30.1%). These two categories make up 69.5% of the total sample. Notice that the gaps between both the second and the third content types and the third and fourth content types are very large. Text is in the third position with 75 stories (17.9%), followed by photo (n=19; 4.5%). Photo & video is at the bottom of this list with only 2 (0.5%) stories. Audio (57.1%), text (32%), and text & photo (25.5%) have their highest percentages in health category; photo & video (100%) and text & video (34.9%) have their highest percentages in religion category; and text & photo & video (61.5%), video (50%), and photo (31.6%) have their highest percentages in the politics category. It is observable that no content types have their highest percentages in entertainment, crime, and miscellaneous categories. Audio (100%) and photo & audio (100%) appeared only in online media. The six other content types have also their highest shares in online media rather than mainstream media: text & video (94.4%) is the highest of them, followed by text & photo & video (92.3%). Of the eight content types, text (17.3%) has the highest percentage in mainstream media, followed by video (16.7%) and text & photo (16.4%).

Two main sources of social media fake news are online media and mainstream media (**Table 3**). Mainstream media mainly includes television channels, newspapers, and radio stations. They are mostly national media outlets. In contrast, online media includes online versions of mainstream television channels and newspapers, online news portals, blogs, various websites, and social media platforms. Of the two, online media (n=366; 87.4%) produces a larger share of fake news than mainstream media (n=53; 12.6%). Of online media, four social media platforms: Twitter, Facebook, YouTube, and WhatsApp are responsible for all fake news. In online media, health-related fake news (n=106; 29%) is on the top of the list, followed by religious fake news (n=92; 25.1%), whereas entertainment-related fake news (n=14; 3.8%) remains on the bottom. In the mainstream media, political fake news (n=19; 35.8%) is the highest, while crime-related fake news (n=1; 1.9%) remains the lowest. If we take fake news contents into account, Table 3 shows that text & photo (n=138; 37.7%) is the dominant content in online media, followed by text & video (n=119; 32.5%); photo & video (n=2; 0.5%) is on the bottom of the list. Like online media, text & photo (n=27; 50.9%) is also the most popular content in mainstream media, followed by text (n=13; 24.5%) with a huge gap in between.

| Table 1: Brief description of themes | | |
|---|---|---|
| Theme | Definition | Example |
| Health | Mainly deals with medicine, medical and healthcare facilities, viral infection, doctor-patient issues, quarantine, and lifestyle. | "Dead bodies in Mecca shared as corona victims", "Medicine will be sprayed in the air to kill coronavirus". |
| Religion | Includes both religious and religiopolitical (a combination of religion and politics) news, dealing with spirituality, practices, and divinity, religious policy, and communalism. | "Trump offers Islamic prayers amid corona", "Muslims are being buried alive in India". |
| Politics | Related to institutional politics, political issues, and political figures. | "Kejriwal admits of having family ties with RSS", "Rahul Gandhi blames PM Modi". |
| Crime | Related to killing, violence, stealing, harassing, and other forms of criminal activity. | "Woman is murdered in Tahir Hussain's house", "Minor girl's death in Madhya Pradesh is linked to Delhi riots". |
| Entertainment | Linked to celebrities and popular culture. | "Salman Khan gifts an apartment to Ranu Mondol", "Korean drama predicted COVID-19". |
| Miscellaneous | Includes the fake news that did not fit in the other five categories, mainly related to military, technology, education, and economy. | "Tata Group of companies will not recruit JNU students", "Mysterious apocalyptic planet spotted in the sky". |

| Table 2: Themes and Contents of Fake News | | | | | | | | | | |
|---|---|---|---|---|---|---|---|---|---|---|
| Theme | | Contents | | | | | | | | |
| | | | Photo | Photo & Video | Text | Text & Photo | Text & Video | Text & Photo & Video | Video | Total |
| **Religion** | Count | 1 | 4 | 2 | 11 | 39 | 44 | 3 | 1 | 105 |
| | % within Themes | 1.0% | 3.8% | 1.9% | 10.5% | 37.1% | 41.9% | 2.9% | 1.0% | 100% |
| | % within Content | 14.3% | 21.1% | 100% | 14.7% | 23.6% | 34.9% | 23.1% | 8.3% | 25.1% |
| **Politics** | Count | 2 | 6 | 0 | 19 | 40 | 21 | 8 | 6 | 102 |
| | % within Themes | 2.0% | 5.9% | 0% | 18.6% | 39.2% | 20.6% | 7.8% | 5.9% | 100% |
| | % within Content | 28.6% | 31.6% | 0% | 25.3% | 24.2% | 16.7% | 61.5% | 50.0% | 24.3% |
| **Miscellaneous** | Count | 0 | 3 | 0 | 7 | 14 | 5 | 1 | 2 | 32 |
| | % within Themes | 0.0% | 9.4% | 0% | 21.9% | 43.8% | 15.6% | 3.1% | 6.3% | 100% |
| | % within Content | 0.0% | 15.8% | 0% | 9.3% | 8.5% | 4.0% | 7.7% | 16.7% | 7.6% |
| **Health** | Count | 4 | 4 | 0 | 24 | 42 | 36 | 1 | 3 | 114 |
| | % within Themes | 3.5% | 3.5% | 0.0% | 21.1% | 36.8% | 31.6% | 0.9% | 2.6% | 100% |
| | % within Content | 57.1% | 21.1% | 0.0% | 32.0% | 25.5% | 28.6% | 7.7% | 25.0% | 27.2% |
| **Entertainment** | Count | 0 | 0 | 0 | 6 | 13 | 2 | 0 | 0 | 21 |

|  | | | | | | | | | |
|---|---|---|---|---|---|---|---|---|---|
| | % within Themes | 0.0% | 0.0% | 0.0% | 28.6% | 61.9% | 9.5% | 0.0% | 0.0% | 100% |
| | % within Content | 0.0% | 0.0% | 0.0% | 8.0% | 7.9% | 1.6% | 0.0% | 0.0% | 5.0% |
| Crime | Count | 0 | 2 | 0 | 8 | 17 | 18 | 0 | 0 | 45 |
| | % within Themes | 0.0% | 4.4% | 0.0% | 17.8% | 37.8% | 40.0% | 0.0% | 0.0% | 100.0% |
| | % within Content | 0.0% | 10.5% | 0.0% | 10.7% | 10.3% | 14.3% | 0.0% | 0.0% | 10.7% |
| Total | Count | 7 | 19 | 2 | 75 | 165 | 126 | 13 | 12 | 419 |
| | % within Themes | 1.7% | 4.5% | 0.5% | 17.9% | 39.4% | 30.1% | 3.1% | 2.9% | 100.0% |
| | % within Content | 100.0% | 100.0% | 100.0% | 100.0% | 100.0% | 100.0% | 100.0% | 100.0% | 100.0% |

Table 3: Sources and Themes of Fake News

| Sources | | Themes | | | | | | |
|---|---|---|---|---|---|---|---|---|
| | | Crime | Entertainment | Health | Miscellaneous | Political | Religious | Total |
| Online Media | Count | 44 | 14 | 106 | 27 | 83 | 92 | 366 |
| | % within Sources | 12.0% | 3.8% | 29.0% | 7.4% | 22.7% | 25.1% | 100.0% |
| | % within Themes | 97.8% | 66.7% | 93.0% | 84.4% | 81.4% | 87.6% | 87.4% |
| Mainstream media | Count | 1 | 7 | 8 | 5 | 19 | 13 | 53 |
| | % within Sources | 1.9% | 13.2% | 15.1% | 9.4% | 35.8% | 24.5% | 100.0% |
| | % within Themes | 2.2% | 33.3% | 7.0% | 15.6% | 18.6% | 12.4% | 12.6% |
| Total | Count | 45 | 21 | 114 | 32 | 102 | 105 | 419 |
| | % within Sources | 10.7% | 5.0% | 27.2% | 7.6% | 24.3% | 25.1% | 100.0% |
| | % within Themes | 100.0% | 100.0% | 100.0% | 100.0% | 100.0% | 100.0% | 100.0% |

ANALYSIS 2: POLICY OPTION FEASIBILITY 'PASTEL' FRAMEWORK

Table 4: Options feasibility "PASTEL" analysis

| | | | | | | |
|---|---|---|---|---|---|---|
| Independent institutional regulation | + | +/- | + | +/- | NA | +/- |
| Joint Self-regulation | + | + | + | + | + | +/- |
| Complete government regulation | + | +/- | - | - | - | + |
| Complete platform regulation | - | +/- | - | NA | + | - |
| Do Nothing | +/- | - | - | NA | + | - |
| Options \ Criteria | Political Feasibility | Administrative Feasibility | Social Feasibility | Technological Feasibility | Economic Feasibility | Legal Feasibility |

Table 5: Stakeholder impact analysis

| | | | | | |
|---|---|---|---|---|---|
| Independent institutional regulation | + | +/- | +/- | +/- | +/- |
| Joint Self-regulation | + | + | +/- | + | +/- |
| Complete government regulation | - | - | + | +/- | - |
| Complete platform regulation | - | - | +/- | - | + |
| Do Nothing | - | + | +/- | - | + |
| Options \ Interest Groups | The Public | Social Media Platforms | The Government | The News Media | Digital Advertisement Firms |

**BIBLIOGRAPHY**

1. Al-Zaman et al., 2020, Social Media Rumors in Bangladesh
   http://koreascience.or.kr/article/JAKO202028851207120.page
2. Humprecht et al., 2020, Resilience to Online Disinformation: A Framework for Cross-National Comparative Research
   https://journals.sagepub.com/doi/10.1177/1940161219900126
3. Nazmi, 2019, Why India shuts down the internet more than any other democracy
   https://www.bbc.com/news/world-asia-india-50819905
4. Burgess, 2018, To fight fake news on WhatsApp, India is turning off the internet
   https://www.wired.co.uk/article/whatsapp-web-internet-shutdown-india-turn-off
5. (Farooq, Politics of Fake News: How WhatsApp became potent propaganda tool in India
   https://www.mediawatchjournal.in/politics-of-fake-news-how-whatsapp-became-a-potent-propaganda-tool-in-india/
6. Joshi, 2020, India's Digital Response to COVID-19 Risks Creating a Crisis of Trust
   https://thewire.in/tech/covid-19-aarogya-setu-surveillance
7. Quandt et al., 2019 Fake News
   https://onlinelibrary.wiley.com/doi/10.1002/9781118841570.iejs0128
8. Allcott & Gentzkow, 2017, Social Media and Fake News in the 2016 Election
   https://www.aeaweb.org/articles?id=10.1257/jep.31.2.211
9. Duffy et al., 2020, Too good to be true, too good not to share: the social utility of fake news
   https://www.tandfonline.com/doi/full/10.1080/1369118X.2019.1623904
10. Muigai, 2019, Understanding Fake News
    http://www.ijsrp.org/research-paper-0119.php?rp=P858104
11. Tandoc et al., 2020, Diffusion of disinformation: How social media users respond to fake news and why
    https://journals.sagepub.com/doi/10.1177/1464884919868325
12. Tandoc et al., 2018, Defining "Fake News"
    https://www.tandfonline.com/doi/full/10.1080/21670811.2017.1360143
13. Netflix, 2020, The Social Dilemma,
    https://www.thesocialdilemma.com/
14. https://www.humanetech.com/
15. Kortina, 2020, The Social Dilemma Dilemma
    https://kortina.nyc/essays/the-social-dilemma-dilemma/#how-social-media-technology-influences-social-behavior
16. Al-Zaman, 2021, Social Media Fake News in India
    https://www.researchgate.net/publication/341725037_Social_Media_Fake_News_in_India
17. Internet Freedom Foundation, supra note 23.
18. Information Technology (Guidelines for Intermediaries and Digital Media Ethics Code) Rules, 2021 (Rules). https://www.businessinsider.in/tech/news/facebook-and-twitter-have-a-ton-of-new-rules-to-abide-by-in-india-whatsapp-may-find-itself-in-the-toughest-spot-of-all/articleshow/81208643.cms
19. https://eci.gov.in/
20. Rasmus Kleis Nielsen, US elections vs Bihar polls: Are all social media users created equal?, Scroll.in (Nov. 10, 2020), https://scroll.in/article/978068/us-elections-vs-bihar-polls-are-all-social-media-users-created-equal.
21. Daniel Funke, In one month, Facebook doubled the countries using its fact-checking tool - all outside the West, Poynter.org (April 18, 2018), https://www.poynter.org/fact-checking/2018/in-one-month-facebook-doubled-the-countries-using-its-fact-checking-tool-%c2%97-all-outside-the-west/, PTI, Facebook teams up with AFP to expand fact-checking programme in India, Business Today (Nov. 6, 2018), https://www.businesstoday.in/current/economy-politics/facebook-partners-with-afp-to-expand-fact-checking-programme-in-india/story/288651.html.

22. PTI, Facebook teams up with 8 third-party fact checkers, covering 11 Indian languages, to flag Covid-19 fake news, The Economic Times (April 22, 2020), https://economictimes.indiatimes.com/magazines/panache/facebook-teams-up-with-8-third-party-fact-checkers-covering-11-indian-languages-to-flag-covid-19-fake news/articleshow/75284845.cms.
23. Gopal Sathe, Fact-Checkers Fight Fake News On Facebook. But Who Fact-Checks Them?, HuffPost (Jan. 1, 2020), https://www.huffpost.com/archive/in/entry/fact-checking-fake-news-facebook-how-does-ifcn-work_in_5ca0fd29e4b00ba6327eb726.
24. Anumeha Chaturvedi, It's up to Facebook to respond on its fact checking programme, say FB fact-checkers after IT minister raises concerns, The Economic Times ET Prime (Sept. 2, 2020), https://economictimes.indiatimes.com/tech/internet/its-up-to-facebook-to-respond-on-its-fact-checking-programme-says-fb-after-it-minister-raises-concerns/articleshow/77895695.cms.
25. Scroll Staff, WhatsApp launches new fact-checking service to fight fake news ahead of elections, Scroll.in (Apr. 02, 2019), https://scroll.in/latest/918725/whatsapp-launches-new-fact-checking-service-to-fight-fake-news-ahead-of-elections.
26. The Wire Staff, To Combat Fake News in India, WhatsApp to Limit Forwarding of Messages, The Wire (July 20, 2018), https://thewire.in/tech/to-combat-fake-news-in-india-whatsapp-to-limit-forwarding-of-messages.
27. Prashant Reddy T., If WhatsApp Doesn't Regulate Itself, Parliament May Have to Step In, The Wire (July 18, 2018), https://thewire.in/tech/if-whatsapp-doesnt-regulate-itself-parliament-may-have-to-step-in.
28. Munsif Vengattil, Aditya Kalra and Sankalp Phartiyal, INSIGHT – In India election, a $14 software tool helps overcome WhatsApp controls, Reuters (May 15, 2019), https://www.reuters.com/article/india-election-socialmedia-whatsapp-idINL4N22R3G3.
29. iPleaders, 2021, How can India regulate misinformation on social media during COVID-19 https://blog.ipleaders.in/can-india-regulate-misinformation-social-media-covid-19/
30. Economic Times, 2022, ETtech Opinion: It's time for a fresh look at social media regulations https://economictimes.indiatimes.com/tech/catalysts/ettech-opinion-its-time-for-a-fresh-look-at-social-media-regulations/articleshow/89724789.cms
31. Cusumano, HBR, 2021, Social Media Companies Should Self-Regulate. Now. https://hbr.org/2021/01/social-media-companies-should-self-regulate-now
32. Varun Thomas Mathew, The Arrogance of Being Facebook, a Serious Tragedy for the Rule of Law, The Wire (Oct. 8, 2020) https://thewire.in/law/facebook-delhi-assembly-summons-rule-of-law-riots.
33. Megha Mandavia, Social Media to Join Hands to Fight Fake News, Hate Speech, The Economic Times (Feb. 19, 2020), https://economictimes.indiatimes.com/tech/internet/social-media-to-join-....
34. Id.
35. Id.
36. European Parliament, Disinformation and propaganda – impact on the functioning of the rule of law in the EU and its Member States (Feb. 2019) at p. 104, https://www.europarl.europa.eu/RegData/etudes/STUD/2019/608864/IPOL_STU(2019)608864_EN.pdf, See also Oxford Handbook of Regulation 541 (Robert Baldwin, Martin Cave, and Martin Lodge eds., 2010).

37. d.; Article 19, The Social Media Councils: Consultation Paper (June, 2019) at p. 7, https://www.article19.org/wp-content/uploads/2019/06/A19-SMC-Consultation-paper-2019-v05.pdf.
38. Id.
39. Olivier Sylvain, Intermediary Design Duties, 50 Conn. L. Rev. 203 (2018), https://ir.lawnet.fordham.edu/cgi/viewcontent.cgi?article=1892&context=f....
40. See generally EU Code of Practice on Disinformation, https://ec.europa.eu/digital-single-market/en/news/code-practice-disinformation (last visited Nov. 20, 2020), Daphne Keller, Who Do You Sue? - State and Platform Hybrid Power Over Online Speech , Hoover Institution, Aegis Series Paper No. 1902 https://www.hoover.org/sites/default/files/research/docs/who-do-you-sue-state-and-platform-hybrid-power-over-online-speech_0.pdf
41. "Misinformation and news quality on digital platforms in Australia - A position paper to guide code development", Australian Communications and Media Authority (June 2020), https://www.acma.gov.au/sites/default/files/2020-06/Misinformation%20and%20news%20quality%20position%20paper.pdf.
42. [42] Id. at p. 11, These risks or acute harms should include— health and safety of vulnerable groups, threat of harm to public and private property, threat to elections or democratic processes, and imminent financial harm.
43. Id. at. p.1.
44. Id. at. p.31-32.
45. Id.
46. Id.
47. Id.
48. Fighting Fake News – Workshop Report, Information Society Project (Mar. 17, 2017), p. 7, https://law.yale.edu/sites/default/files/area/center/isp/documents/fighting_fake_news_-_workshop_report.pdf.
49. Gesetz zur Verbesserung der Rechtsdurchsetzung in sozialen Netzwerken [Netzwerkdurchsetzungsgesetz—NetzDG] [Network Enforcement Act], Sept. 1, 2017, Bundesgesetzblatt, Teil I [BGBl I] at 3352 (Ger.).
50. Protection from Online Falsehoods and Manipulation Act, 2019, https://sso.agc.gov.sg/Acts-Supp/182019/Published/20190625?DocDate=20190625#:~:text=An%20Act%20to%20prevent%20the,to%20be%20taken%20to%20enhance (Singapore).
51. It is critical to ensure that the oversight body working with platforms to develop self-regulatory codes is transparent and collaborative. In India especially, a culture of opacity shrouds law-making by state actors and regulators like the Ministry of Electronics and IT are often criticized for the lack of transparency and public participation in framing rules for Internet governance. See, e.g., Ayesha Khan, Submission to the Ministry of Electronics and Information Technology, Government of India, on the Draft Non-Personal Data Governance Framework, Wikimedia/Yale Law School Initiative on Intermediaries and Information Blog (Sept. 16, 2020), https://law.yale.edu/submission-ministry-electronics-and-information-technology-government-india-draft-non-personal-data.
52. Gaur, 2021, Yale law, Towards policy and regulatory approaches for combating misinformation in India https://law.yale.edu/isp/initiatives/wikimedia-initiative-intermediaries-and-information/wiii-blog/towards-policy-and-regulatory-approaches-combating-misinformation-india